\documentclass{article} 
\usepackage{graphicx}        
\begin{document}
\centerline{\bf Marry your Sister: Outbreeding Depression in Penna Ageing Model}
\medskip

\noindent
Stanis{\l}aw Cebrat, Dietrich Stauffer$^1$ 
\medskip

\noindent
Department of Genomics, 
Wroc{\l}aw University, ul. Przybyszewskiego 63/77, 51-148 Wroc{\l}aw, Poland
\medskip

\noindent
$^1$ Visiting from Institute for Theoretical Physics, Cologne University,
D-50923 K\"oln, Euroland
\bigskip

{\small
If in the sexual Penna ageing model conditions are applied leading to 
complementary bit-strings, then marriages between brothers and sisters, or 
between close cousins, may lead to more offspring than for unrelated couples.
}
\medskip

Inbreeding is usually a bad thing for humans and many animals; the Egyptian
pharaos liked marriages between brothers and sisters which led to the
city named Philadelphia but did not prevent conquest by the Romans two
millennia ago. Also in the Penna ageing model \cite{penna}, an often used
computer simulation model for biological ageing, such inbreeding depression
has been found \cite{sousa}: Too small populations die out because their
genomes become too similar, activating many deleterious mutations which in
a larger population with more genomic diversity would remain recessive
and would not affect the health of the individual.

However, outbreeding depression is also possible \cite{bonkowska}: If the
genomes are too different, survival becomes difficult. For example, donkey
and horses cannot have grand children, and the senior author is the only
known {\it homo troglodytes} and has no known children \cite{pmco}. 
This effect may be connected with the possibility that the two haploid
genomes in a species with sexual reproduction are complementary to each 
other \cite{zaw}, where with few exceptions for each gene one haplotype
has one allele (``wild type'') and the other has the opposite allele
(``mutant'') and where most mutations are recessive. Thus, even though 
about half of the genes are mutated, the phenotype is barely affected.

For real humans instead of computer simulations, Helgason et al. \cite{ice}
checked for all known marriages in Iceland 1800-1965 whether they were
cousins and how many offspring they had. They found the greatest reproductive
success, measured in the number of grand children, for third and fourth 
cousins. We now check for similar effects by simulating the Penna ageing model.

Each individual in the Penna model \cite{penna} has a genome of two bit-strings
of length $L$ each; at age $a$ only the first $a$ bit positions are active.
If the two alleles on one position are 00, 01, or 10, they do not affect
the health. If one locus at position $a$ is 11 instead, then the individual
dies at age $a$. For ages above or at a minimum reproduction age $R=5L/8$, 
each surviving female at each iteration randomly tries to find a male of 
reproductive age and then has $B$ children with him; the two bit-strings within
each parent are copied and the copies are crossed-over randomly with a 
probability $C$ and suffer from one 
deleterious mutation in each bit-string. Thus the mutations are recessive,
irreversible, and lethal if present in both alleles. The population size 
fluctuates about a value normally somewhat below the carrying capacity
$K$ (Verhulst death probability, applied to babies only).  20,000 or 40,0000
iterations (updates of each survivor) are made for equilibration; then for 1400 
iterations we analyse for cousins and their offspring.

For the first half of these 1400 analysis iterations, we check at each
marriage five generations back for the first ancestor which is common to
husband and wife. Agreement in the $n$-th generation back gives a genealogical
distance of $n$; if husband and wife have different $n$, the larger of the two 
is taken as genealogical distance; this happens if on old man marries
his niece, etc. If no common ancestor is found within the investigated
five generations back, the genealogical distance is set to 6. Thus second-order
cousins have genealogical distance 3.
 
During the second half (again 700 iterations) of the analysis, we also check 
how many children (level=1), grand children (level=2) etc, again up to five 
generations (level=5), couples of genealogical distance $n$ have given birth to.
In this way we get a matrix with the two variables level and genealogical 
distance. (If the levels of the two parents differ, the son adds 1 to the 
paternal level, and the daughter adds 1 to the maternal level.)

\begin{figure}[hbt]
\begin{center}
\includegraphics[angle=-90,scale=0.5]{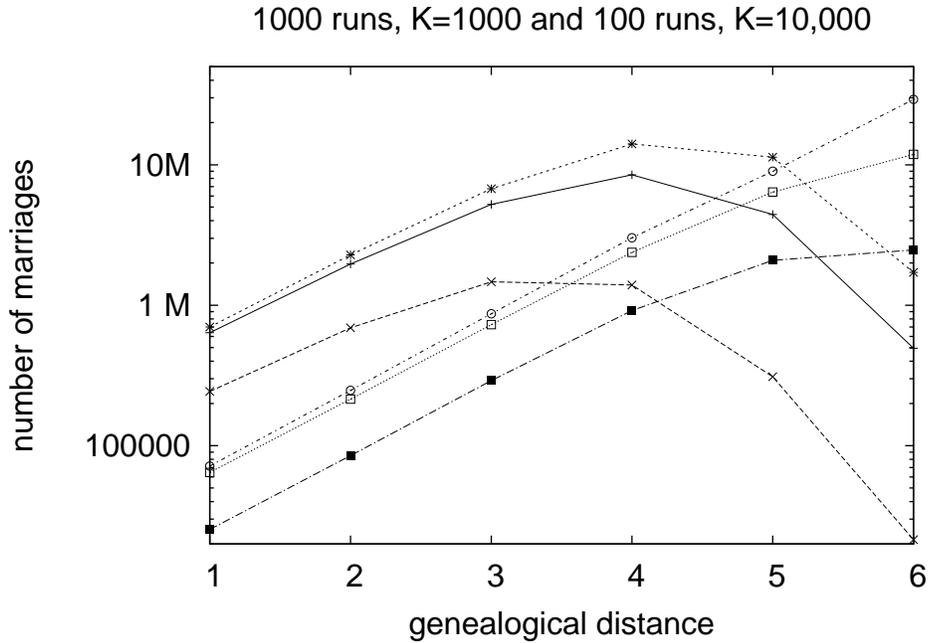}
\end{center}
\caption{Number of marriages versus genealogical distance, for $K = 1000$
(curves with maximum) and 10,000 (increasing curves). The recombination rates 
are $C=0.001$ (+ and open squares), 1 (stars and circles) and $\times$ 
($C=0.128, \, K=1000$) and full squares ($C=0.032$ at $K=10,000$). 
}
\end{figure}

In a small population, our random selection of partners automatically leads
to marriages between close cousins, and relatively few couples have a 
genealogical distance larger than 5. In a large population, few couples
are close cousins, and most have a distance larger than 5. Thus as a function
of distance $n$, a maximum in the number of marriages is expected, shifting
to larger $n$ for larger carrying capacities $K$. Fig.1 shows this effect:
For small $K = 1000$ the curves show the maximum, for large $K=10,000$ the 
maximum is beyond $n=6$. We took $C= 0.001$ and 1, as well as an intermediate
$C$ close to the complementarity transition.
 
In general we found little dependence of the number of offspring
on the genealogical distance for large 
recombination rates $C$ where the genomic bit-strings are not complementary.
With decreasing $C$ the surviving population first reaches a minimum or dies out
and then increases again for lower $C$, with complementary bit-strings (as 
shown by analysis of bits set to one and of heterozygous loci; not shown). 
Then close cousins usually have somewhat more offspring than unrelated couples. 
Fig.2 shows results per marriage for $L=64, \, B=2, \, K=1000$ at $C = 0.032$; 
most of the populations died out at $C = 0.128$ 
($C$ is increased from 0.001 by factors of two).
Each figure part shows the averages over 100 (+), 1000 ($\times$) and 10,000 (*)
samples. Thus the stars should be relied upon; their distance from the plus 
signs is ten times their statistical error. The results are roughly independent
of the offspring level. This decrease with increasing genealogical
distance also is seen near the complementarity transition, also for larger 
$L=128$ and smaller $L=32$: Fig.3. The final age distribution showed little
dependence on the parental genealogical distance (not shown). 

The reproductive fraction is the fraction of survivors at the end of the 
simulation with an age at or above the minimum reproduction age. Figure 
4 shows that it depends little on the genealogical distance $n$ of the
parents or the crossing rate $C$, except that near the transition to 
complementarity an upturn is seen towards $n=6$ (meaning again all
distances larger than 5).

In summary, we found decreasing numbers of offspring with increasing 
genealogical distance $n$ at low $C$, and little influence of $n$ at large $C$. 
Reality \cite{ice} with a maximum at $n \simeq 4$ seems to be in between.

\begin{figure}[hbt]
\begin{center}
\includegraphics[angle=-90,scale=0.32]{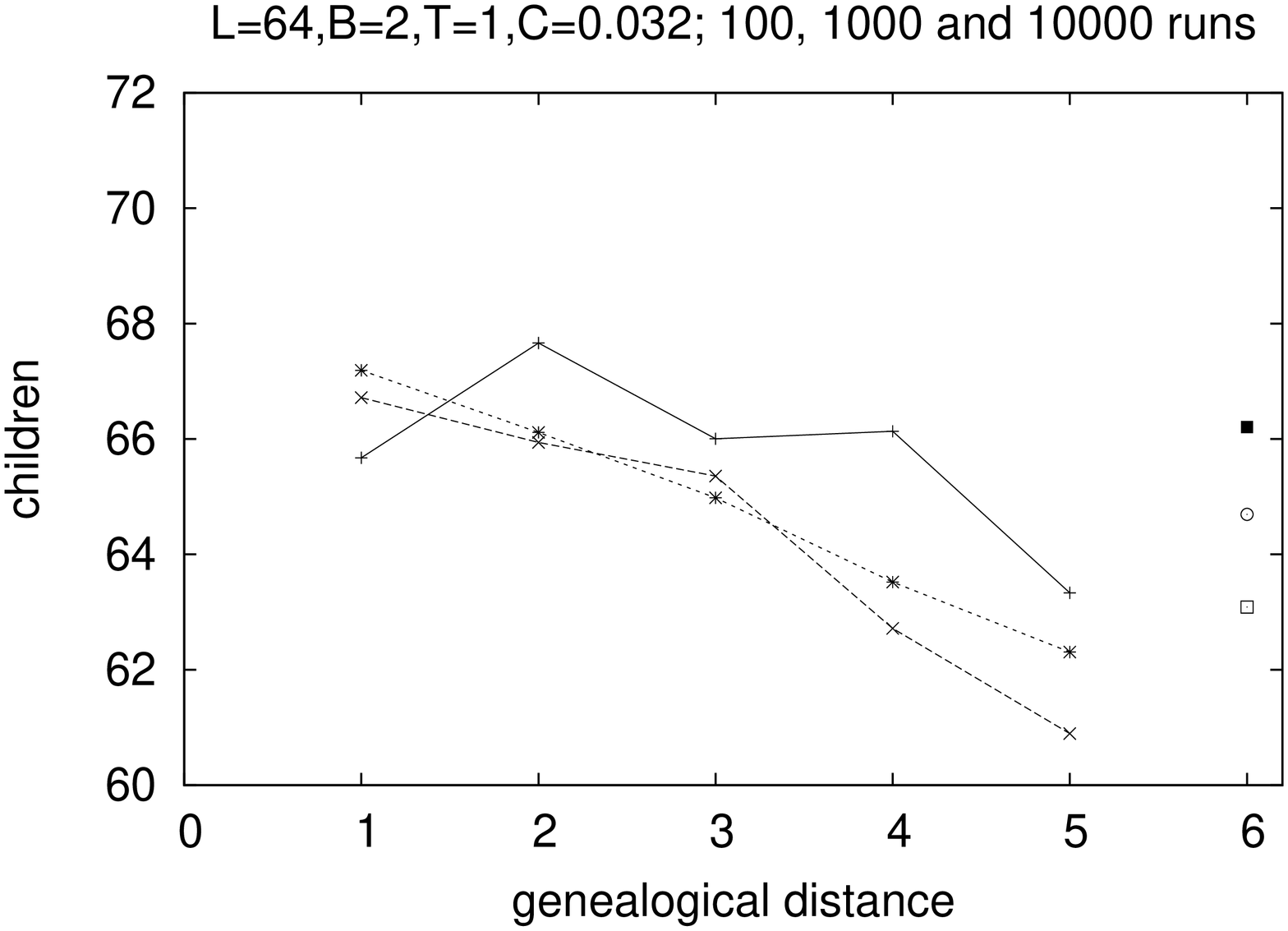}
\includegraphics[angle=-90,scale=0.32]{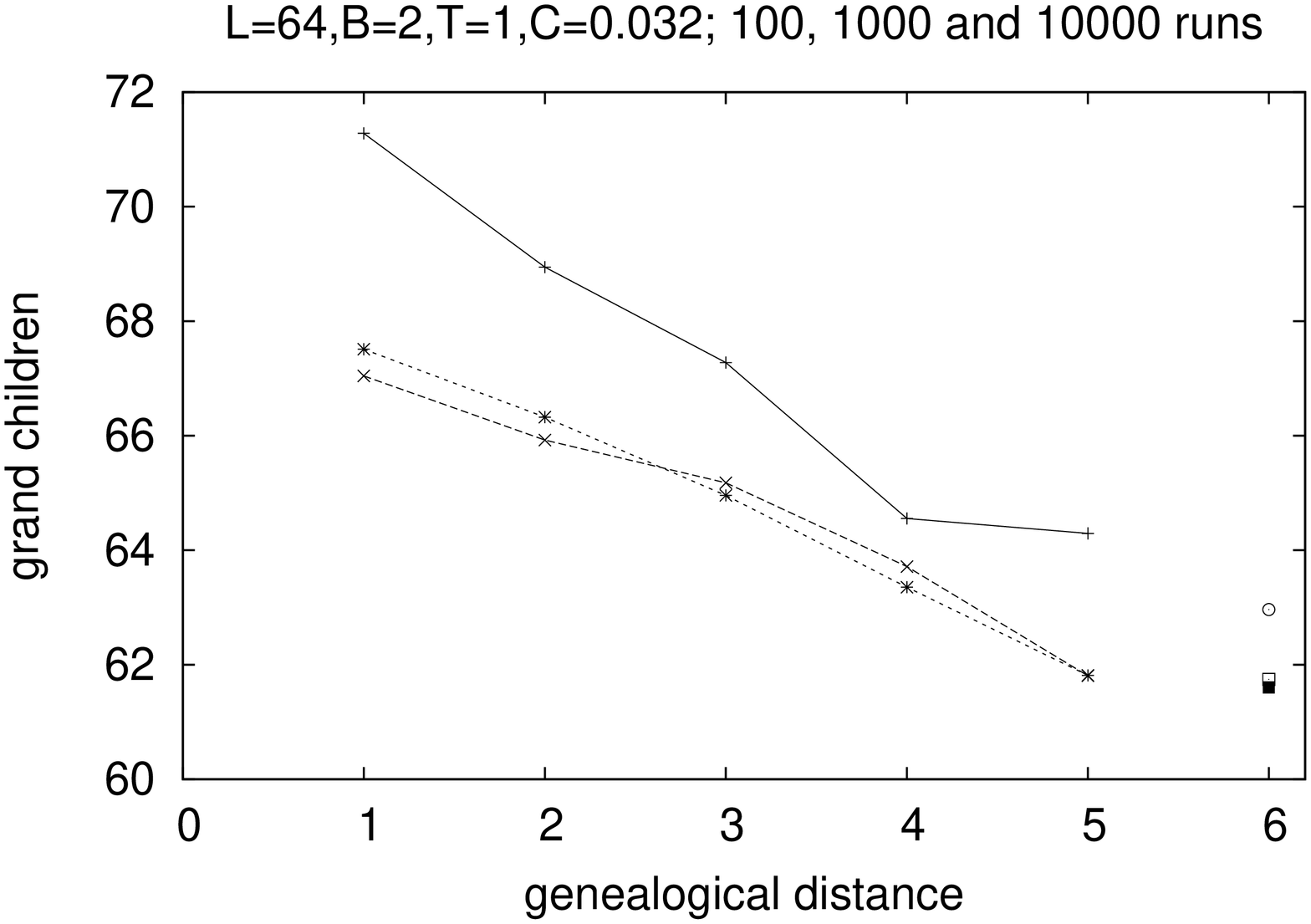}
\includegraphics[angle=-90,scale=0.32]{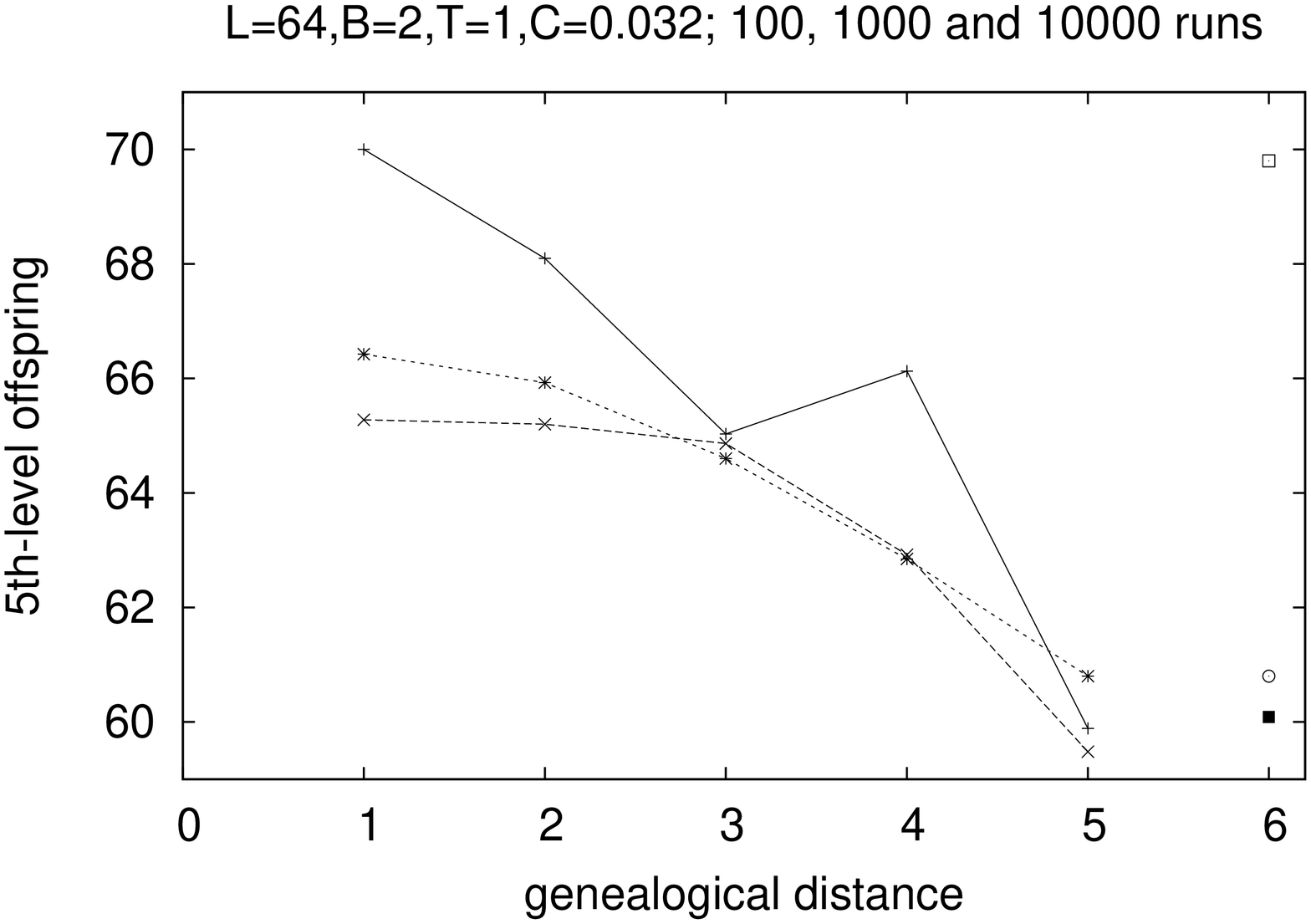}
\end{center}
\caption{Number of children (top part), grand children (central part), and
5th-level offspring (bottom part) versus genealogical distance $n$ where the
separate symbols at $n=6$ refers to couples without known relation ($n > 5$). 
Averages over 100 (+, open squares), 1000 ($\times$, full squares) and 10,000 
(stars, circles) samples at $K=1000$. 
}
\end{figure}

\begin{figure}[hbt]
\begin{center}
\includegraphics[angle=-90,scale=0.4]{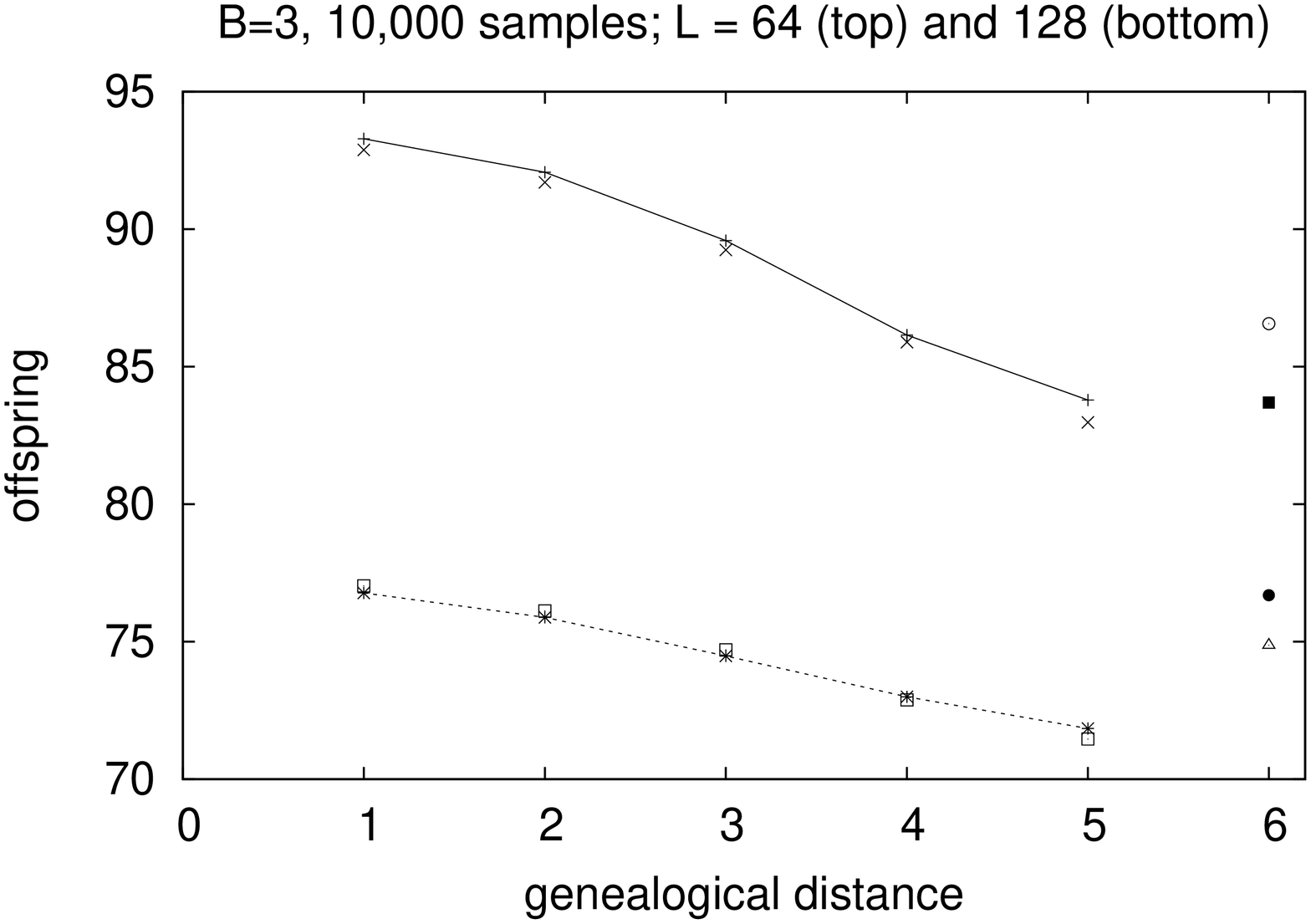}
\includegraphics[angle=-90,scale=0.4]{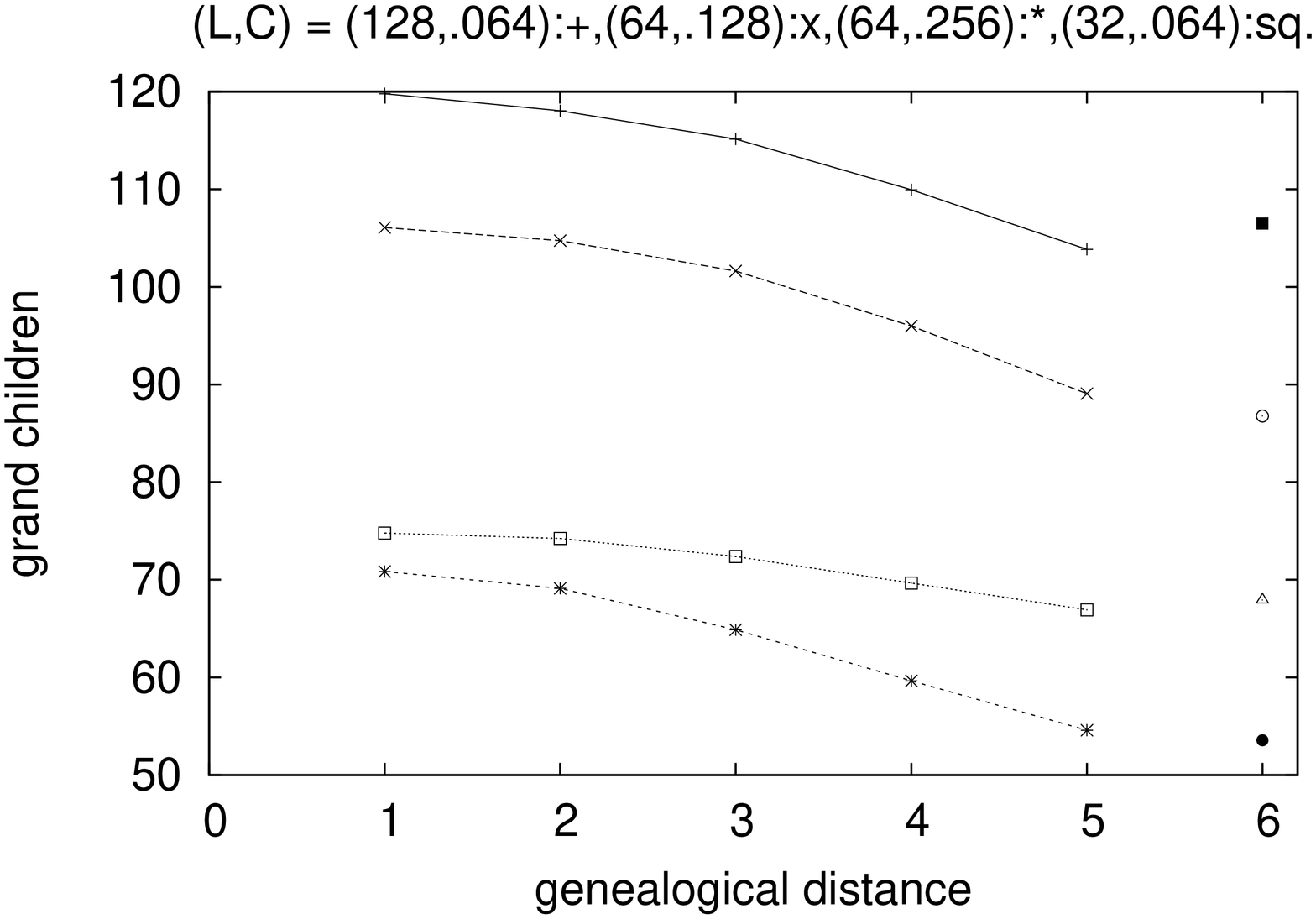}
\end{center}
\caption{Top part: Number of children ($+,*$) and grand children ($\times$,open 
squares) offspring for $L=64,\,C=0.128$ (upper data) and for $L=128,\,C=0.064$ 
(lower data), all at $B=3$ near the complementarity transition with 10,000 runs.
Bottom part: grand children from 10,000 samples for $L=128, \, C=0.064 
(+); \;\; L=64, \, C=0.128 (\times); \;\; L=64, \, C=0.256 (*)$, for $B=4$;
in the last case no complementarity was seen. Finally the squares represent
$L=32, \, C=0.064, \, B=8$. 
}
\end{figure}

\begin{figure}[hbt]
\begin{center}
\includegraphics[angle=-90,scale=0.4]{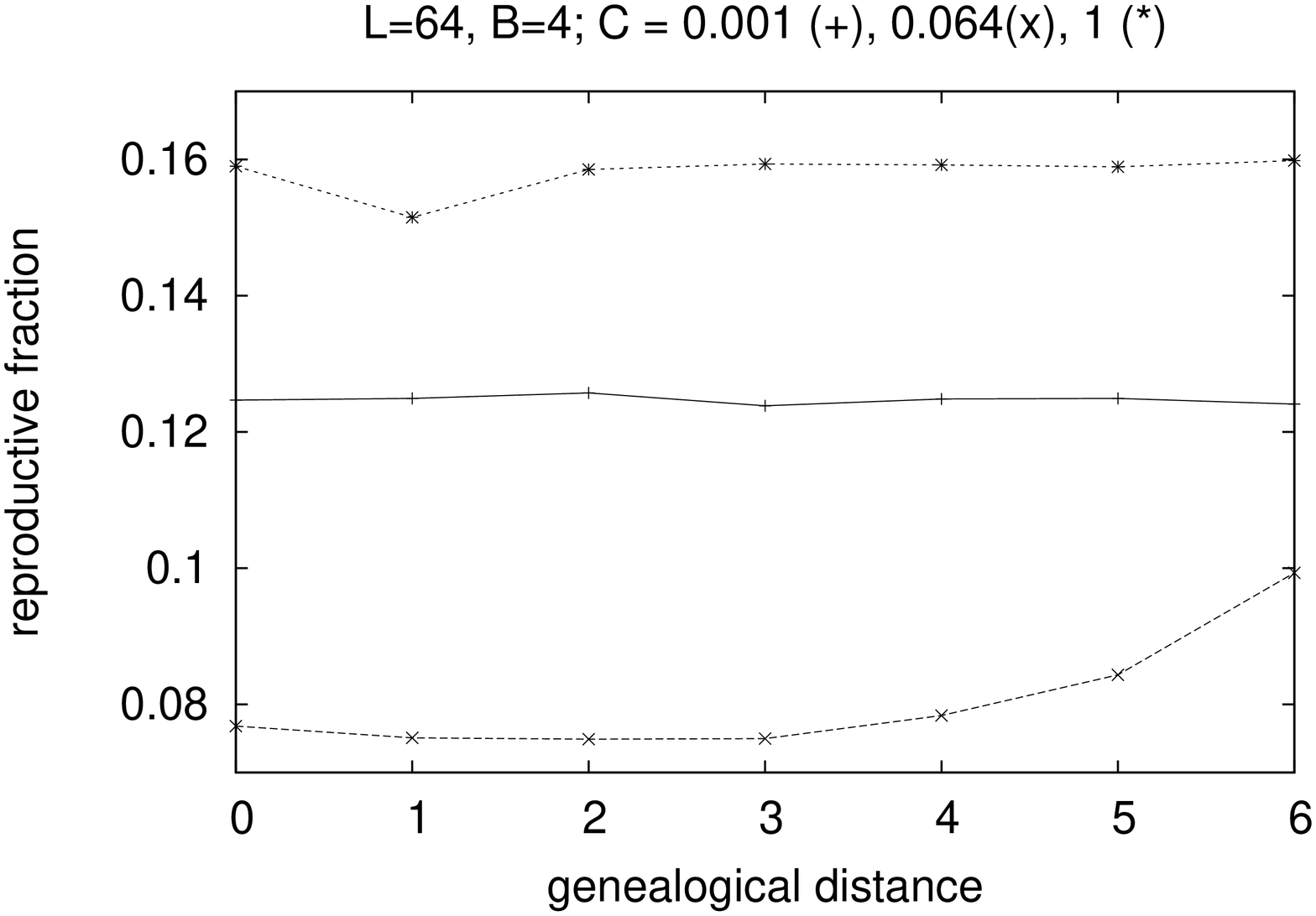}
\includegraphics[angle=-90,scale=0.4]{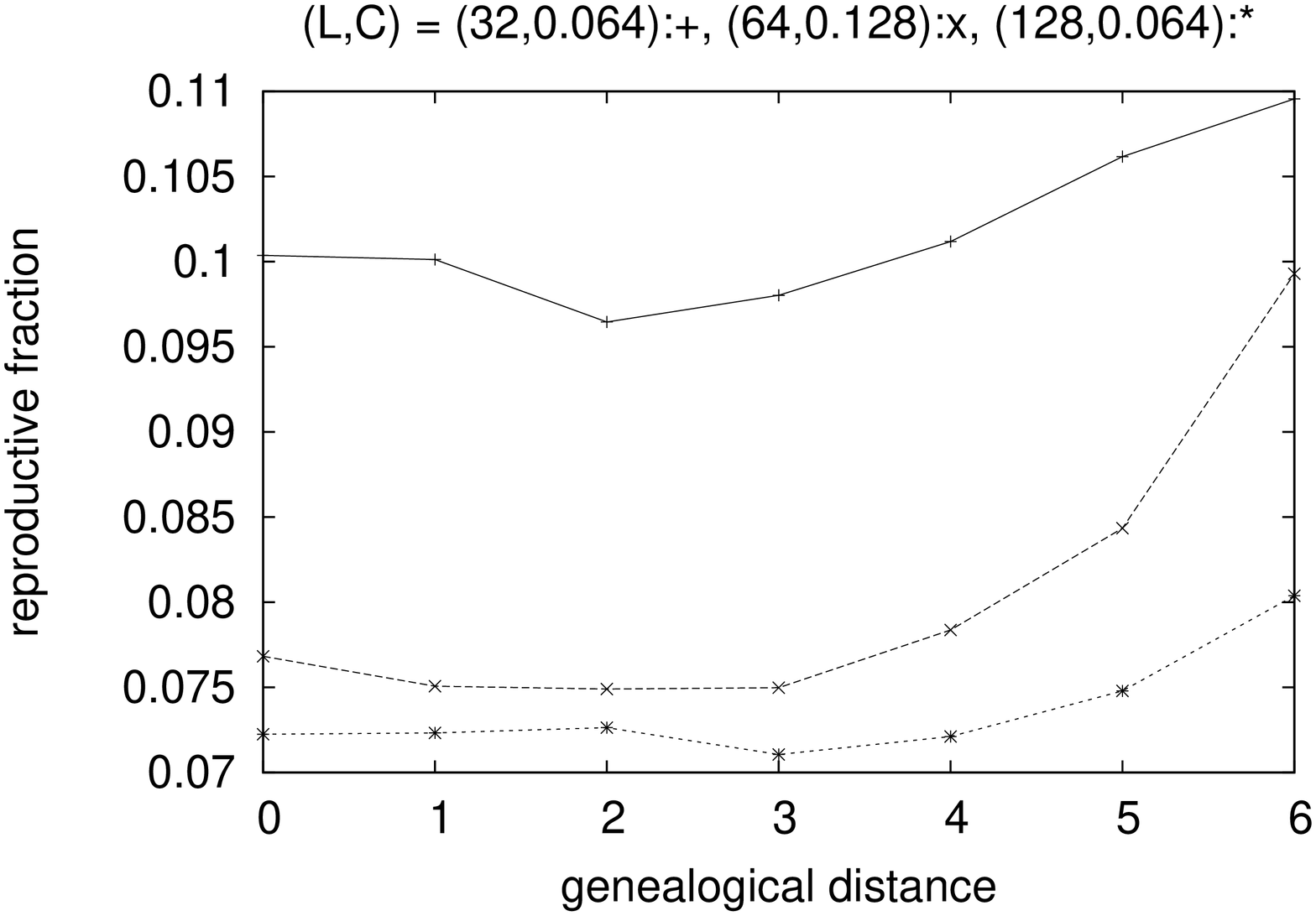}
\end{center}
\caption{Reproductive fractions from 10,000 samples at $K=1000$, where the sum 
over all genealogical
distances is plotted at zero, and that over all distances above 5 at 6: 
Top part: $L=64, \, B=4$ at: $C = 0.001 (+), \, 0.128 (\times), 
\; 1 (*)$. Bottom part near the transition to complementarity at:
$L = 32, \, B=8, \, C=0.064 (+); \; L= 64, \, B=4, \, C=0.128 (\times); \; 
L=128, \, B=4, \, C=0.064 (*)$. 
}
\end{figure}

\end{document}